\documentclass[11pt]{article}

\usepackage{amssymb,amsthm,amscd,array}
\usepackage{graphics}
\usepackage[final]{epsfig}

\let\ssection=\section
\renewcommand{\section}{\setcounter{equation}{0}\ssection}

\newcommand{\bbR}{\mathbb{R}}

\newcommand{\bbZ}{\mathbb{Z}}

\newcommand{\bbC}{\mathbb{C}}
\newcommand{\bbK}{\mathbb{K}}

\newcommand{\ad}{\mathrm{ad}}
\newcommand{\asl}{\mathrm{asl}}
\newcommand{\ah}{\mathrm{ah}}
\newcommand{\cc}{\Gamma}
\newcommand{\cB}{{\mathcal{B}}}
\newcommand{\E}{\mathcal{E}}
\newcommand{\cE}{\mathcal{E}}

\newcommand{\id}{\textup{Id}}

\newcommand{\cK}{{\mathcal{K}}}
\newcommand{\cAK}{{\mathcal{AK}}}

\newcommand{\cP}{{\mathcal{P}}}

\newcommand{\cF}{{\mathcal{F}}}

\newcommand{\End}{\mathrm{End}}
\newcommand{\Der}{\mathrm{Der}}

\newcommand{\rE}{\mathrm{E}}

\newcommand{\osp}{\mathrm{osp}}
\newcommand{\OSp}{\mathrm{OSp}}

\newcommand{\half}{\frac{1}{2}}
\newcommand{\fg}{\mathfrak{g}}

\newcommand{\fa}{\mathfrak{a}}

\chardef\s=110
\chardef\g=103

\begin{document}

\font\rus=wncyr10 scaled 1000

\newtheorem{theorem}{Theorem}%[section]
\newtheorem{lemma}{Lemma}[section]
\newtheorem{cor}[lemma]{Corollary}
\newtheorem{conj}[lemma]{Conjecture}
\newtheorem{proposition}[lemma]{Proposition}
\newtheorem{rmk}[lemma]{Remark}
\newtheorem{exe}[lemma]{Example}
\newtheorem{defi}[lemma]{Definition}

\def\a{\alpha}
\def\b{\beta}
\def\d{\delta}
\def\e{\varepsilon}
\def\i{\iota}
\def\g{\gamma}
\def\L{\Lambda}
\def\om{\omega}
\def\r{\rho}
\def\s{\sigma}
\def\t{\tau}
\def\vfi{\varphi}
\def\vr{\varrho}
\def\l{\lambda}
\def\m{\mu}

\title{Lie antialgebras: cohomology and representations}

\author{V. Ovsienko}

\date{}

\maketitle

\begin{abstract}
We describe the main algebraic and geometric properties of
the class of algebras introduced in \cite{Ovs1}.
We discuss their origins in symplectic geometry and associative algebra,
and the notions of cohomology and representations.
We formulate classification theorems and give a number of examples.
\end{abstract}

\thispagestyle{empty}

%\tableofcontents

%%%%%%%%%%%%%%%%%%%%%%%%%%%%%%%%%%%%%%%%%%%%%
%%%%%%%%%%%%%%%%%%%%%%%%%%%%%%%%%%%%%%%%%%%%%
\section{Introduction}
%%%%%%%%%%%%%%%%%%%%%%%%%%%%%%%%%%%%%%%%%%%%%
%%%%%%%%%%%%%%%%%%%%%%%%%%%%%%%%%%%%%%%%%%%%%

Lie antialgebras is a new class of algebras with origins in 
symplectic and contact geometry of $\bbZ_2$-graded space \cite{Ovs1}. 
These algebras also have a very simple algebraic definition that consists in a (partial)
skew-symmetrization of the associativity condition \cite{LO}.
The most interesting example of Lie antialgebra
is the conformal Lie antialgebra which is related to conformal field theory
(cf. \cite{Gie,GO}).

This paper is a survey of the results obtained in \cite{GO,Ovs1,LO,MG}.
We pay a special attention to the notions of representations and cohomology
of Lie antialgebras and discuss the relations to Lie superalgebras.
We present the detailed explanation of the main definitions and examples,
but we omit the technical proofs.

%%%%%%%%%%%%%%%%%%%%%%%%%%%%%%%%%%%%%%%%%%%%%
\subsection{The definition}
%%%%%%%%%%%%%%%%%%%%%%%%%%%%%%%%%%%%%%%%%%%%%

The following definition is equivalent to the original definition of \cite{Ovs1}.

\begin{defi}
\label{GhostAl}
{\rm
A Lie antialgebra is a $\bbZ_2$-graded vector space $\fa=\fa_0\oplus\fa_1$
equipped with a bilinear product $].,.[$ preserving the parity and
satisfying the following properties:

\begin{enumerate}

\item
supercommutativity, i.e., for homogeneous elements $x,y\in\fa$
\begin{equation}
\label{SkewP}
]x,y[\,=\,(-1)^{p(x)p(y)}
\,]y,x[,
\end{equation}
where $p$ is the parity function: $p|_{\fa_0}=0$ and $p|_{\fa_1}=1$;

\item
$\fa_0$ is a commutative associative subalgebra, one has:
\begin{equation}
\label{AssCommT}
\left]x_1,\left]x_2,x_3\right[\right[
\,=\,
\left]\left]x_1,x_2\right[,x_3\right[
\qquad
\hbox{for all}
\quad
x_i\in{}\fa_0\,;
\end{equation}

\item
for all $x_1,x_2\in{}\fa_0$ and $y\in{}\fa_1$, one has:
\begin{equation}
\label{CacT}
\textstyle
\left]x_1,
\left]x_2,y
\right[
\right[
=
\half
\left]
\left]x_1,x_2
\right[,y
\right[\,;
\end{equation}

\item
for all $x\in{}\fa_0$ and $y_1,y_2\in{}\fa_1$
the following Leibniz rule is satisfied:
\begin{equation}
\label{ICommT}
\left]x,\left]y_1,y_2\right[
\right[=
\left]
\left]x,y_1
\right[,y_2
\right[+
\left]y_1,
\left]x,y_2
\right[
\right[\,;
\end{equation}

\item
for all $y_i\in{}\fa_1$, one has the Jacobi-type identity:
\begin{equation}
\label{Jack}
\left]y_1,
\left]y_2,y_3
\right[
\right[+
\left]y_2,
\left]y_3,y_1
\right[
\right[+
\left]y_3,
\left]y_1,y_2
\right[
\right[=0.
\end{equation}
\end{enumerate}
}
\end{defi}

%%%%%%%%%%%%%%%%%%%%%%%%%%%%%%%%%%%%%%%%%%%%%
\subsection{Examples}\label{DE}
%%%%%%%%%%%%%%%%%%%%%%%%%%%%%%%%%%%%%%%%%%%%%

\begin{exe}
{\rm
The first example of a Lie antialgebra (found in \cite{GO}) is the
\textit{simple} infinite-dimensional algebra with the basis 
$\{e_n,\,n\in\bbZ;\,\ell_m\,m\in\bbZ+\half\}$
and the following relations:
\begin{equation}
\label{GhosRel}
\begin{array}{rcl}
\left]e_n,e_m\right[ &=&
e_{n+m},\\[8pt]
\left]e_n,\ell_i\right[ &=&
\half\,\ell_{n+i},\\[8pt]
\left]\ell_i,\ell_j\right[ &=&
\half\left(i-j\right)e_{i+j}.
\end{array}
\end{equation}
This algebra is called the \textit{conformal Lie antialgebra},
it is denoted by $\cAK(1)$.
The elements $e_n$ span the even part $\cAK(1)_0$, while 
the elements $\ell_m$
span the odd part $\cAK(1)_1$.
}
\end{exe}

The conformal Lie antialgebra $\cAK(1)$ is related to the
classical \textit{conformal Neveu-Schwarz Lie superalgebra}, $\cK(1)$.
Recall that $\cK(1)$ has the basis
$
\textstyle
\left\{
x_n,\;n\in\bbZ
\,;
\xi_i,\;i\in\bbZ+\half
\right\}
$
subject to the following commutation relations
\begin{equation}
\label{CAlgRel}
\begin{array}{rcl}
\left[
x_i,x_j
\right] &=&
\left(j-i\right)x_{i+j},\\[8pt]
\left[
x_i,\xi_j
\right] &=&
\left(j-\frac{i}{2}\right)\xi_{i+j},\\[8pt]
\left[
\xi_i,\xi_j
\right] &=&
2\,x_{i+j}.
\end{array}
\end{equation}
The following result was obtained in \cite{Ovs1}.

\begin{theorem}
\label{DerThm}
The conformal algebra $\cK(1)$ is the algebra of symmetry
of $\cAK(1)$:
$$
\cK(1)=\Der
\left(
\cAK(1)
\right).
$$
\end{theorem}

\noindent
More precisely, the action of $\cK(1)$ on
$\cAK(1)$ is given by
\begin{equation}
\label{CAactRel}
\begin{array}{rclrcl}
x_n(e_m) &=&
m\,e_{n+m},&
\xi_i(e_n) &=&
\ell_{i+n}.
\\[8pt]
x_n(\ell_i) &=&
\left(i-\frac{n}{2}\right)\ell_{n+i},&
\xi_i(\ell_j) &=&
\left(j-i\right)e_{i+j},
\end{array}
\end{equation}
this action preserves the operation (\ref{GhosRel}).

\begin{exe}
{\rm
The most elementary non-trivial example of a Lie antialgebra is
of dimension~$1|2$.
It is spanned by the elements $\{\e;\,a,b\}$
subject to the following relations:
$$
\textstyle
\left]\e,\e\right[=\e,
\qquad
\left]\e,a\right[=\half\,a,
\qquad
\left]\e,b\right[=\half\,b,
\qquad
\left]a,b\right[=\half\,\e.
$$
This is a simple Lie antialgebra denoted by $\asl(2)$.
Note that, for every $i\in\bbZ+\half$, the elements $\{e_0;\,\ell_i,\ell_{-i}\}$
of $\cAK(1)$ generates a subalgebra isomorphic to $\asl(2)$.
The Lie antialgebra $\asl(2)$ is related to the classical simple Lie superalgebra
$\osp(1|2)$, namely
$
\Der(\asl(2))
\cong
\osp(1|2).
$
}
\end{exe}

The classification of simple finite-dimensional Lie antialgebras over
$\bbK=\bbC$ or $\bbR$ was obtained in \cite{Ovs1}.
The result is precisely the same as for commutative algebras.

\begin{theorem}
(i) $\asl(2,\bbC)$ is the only complex simple
finite-dimensional Lie antialgebra; 

(ii) there are two real simple finite-dimensional Lie antialgebras: 
$\asl(2,\bbR)$ and $\asl(2,\bbC)$ (viewed as a $4|2$-dimensional real Lie
antialgebra).
\end{theorem}

A number of examples of finite-dimensional Lie antialgebras
is given in \cite{Ovs1}, let us mention here one of them.

\begin{exe}
{\rm
Consider a family of Lie antialgebras of
dimension $1|2$ with basis $\{\a;\,a,b\}$ and
the relations
$$
\textstyle
\left]\a,\a\right[=0,
\qquad
\left]\a,a\right[=\kappa\,b,
\qquad
\left]\a,b\right[=0,
\qquad
\left]a,b\right[=\half\,\a,
$$
where $\kappa$ is a constant.
If $\kappa=0$, then we call it the Heisenberg antialgebra $\ah_1$,
if $\kappa\not=0$, then the defined
Lie antialgebra is a non-trivial
deformation of $\ah_1$.
}
\end{exe}

%%%%%%%%%%%%%%%%%%%%%%%%%%%%%%%%%%%%%%
%%%%%%%%%%%%%%%%%%%%%%%%%%%%%%%%%%%%%%
\section{Origins of Lie antialgebras}
%%%%%%%%%%%%%%%%%%%%%%%%%%%%%%%%%%%%%%
%%%%%%%%%%%%%%%%%%%%%%%%%%%%%%%%%%%%%%

We present two different ways to obtain Lie antialgebras.
The first one is geometric \cite{Ovs1}, the second one is purely
algebraic \cite{LO}.

%%%%%%%%%%%%%%%%%%%%%%%%%%%%%%%%%%%%%%
\subsection{Lie antialgebras and symplectic geometry}
%%%%%%%%%%%%%%%%%%%%%%%%%%%%%%%%%%%%%%

Lie antialgebras first appeared in symplectic geometry.

Consider the space $\bbR^{2|1}$ equipped with
the standard symplectic form
$$
\textstyle
\om=dp\wedge{}dq+\half\,d\t\wedge{}d\t,
$$
where $p$ and $q$ are even coordinates
and $\t$ is an odd coordinate,
so that $\t^2=0$.
The Poisson bivector field $\cP=\om^{-1}$ is given by
$$
\cP=\frac{\partial}{\partial{}p}
\wedge\frac{\partial}{\partial{}q}+
\half\,
\frac{\partial}{\partial{}\t}
\wedge\frac{\partial}{\partial{}\t}.
$$
The Lie supergroup of linear symplectic transformations 
is denoted by $\OSp(1|2)$ and the corresponding Lie superalgebra
by $\osp(1|2)$.

The bivector field $\cP$ is, of course, $\OSp(1|2)$-invariant.
A remarkable fact is that there exists another, \textit{odd},
invariant bivector field:
\begin{equation}
\label{LaMb}
\Lambda=
\frac{\partial}{\partial\t}\wedge\cE+
\t\,\frac{\partial}{\partial{}p}
\wedge\frac{\partial}{\partial{}q},
\end{equation}
where
$$
\cE=p\,\frac{\partial}{\partial{}p}+
q\,\frac{\partial}{\partial{}q}+
\t\,\frac{\partial}{\partial\t}
$$
is the Euler vector field.
To the best of our knowledge, the bivector field (\ref{LaMb})
was overlooked for years (see however \cite{Gie, GLS1} for its non-explicit forms)
and first explicitly written in \cite{GO}.
The following statement is proved in \cite{Ovs1}.

\begin{theorem}
\label{IntLam}
Every $\OSp(1|2)$-invariant bivector field on $\bbR^2$
is a linear combination of $\cP$ and $\Lambda$.
\end{theorem}

The bivector field (\ref{LaMb}) defines an \textit{odd} bilinear operation on the space
of functions on $\bbR^{2|1}$
\begin{equation}
\label{AntiPo}
\left]F,G\right[=
\frac{(-1)^{p(F)}}{2}\,
\langle
\Lambda,dF\wedge{}dG
\rangle.
\end{equation}
Since $\Lambda$ is with linear coefficients, the space of linear
functions with the bracket (\ref{AntiPo}) form an algebra.
It turns out that \textit{this algebra is
isomorphic to} $\asl(2,\bbK)$.

The full space of functions on $\bbR^{2|1}$
is \textit{not} a Lie antialgebra.
We consider the space, $\cF_\l$, of
\textit{homogeneous} functions of degree $\l$,
i.e., satisfying the condition
$$
\cE(F)=\l\,F,
\qquad\l\in\bbR
$$
(we allow singularities at $p=0$).
The bracket (\ref{AntiPo}) defines a bilinear map
$$
].,.[:\cF_\l\otimes\cF_\mu\to\cF_{\l+\mu-1},
$$
for instance, the space $\cF_1$ is a subalgebra.
Choose the following ``basis'' which is dense in $\cF_1$:
\begin{equation}
\label{Taylor}
\textstyle
\ell_i=p\,\left(\frac{q}{p}\right)^{i+\half},
\qquad
e_n=\t\left(\frac{q}{p}\right)^n
\end{equation}
and substitute it to the bracket (\ref{AntiPo}).

\begin{proposition}
\label{HomAntilie}
The functions (\ref{Taylor}) span the conformal Lie antialgebra $\cAK(1)$.
\end{proposition}

\noindent
Note that the even functions $\ell_i$ in (\ref{Taylor}) span the odd
part $\cAK(1)_1$ while the odd functions $e_n$ span $\cAK(1)_0$.
This is due to the fact that the bivector $\Lambda$ is odd,
and this is the reason of the term ``antialgebra'' that we use.

%%%%%%%%%%%%%%%%%%%%%%%%%%%%%%%%%%%%%%
\subsection{Algebraic definition}
%%%%%%%%%%%%%%%%%%%%%%%%%%%%%%%%%%%%%%

Given a vector space $E$ and a bilinear map
$m:E\times{}E\to{}E$,
the ``associator''
(also called the Gerstenhaber product of $m$ with itself,
see \cite{Ger}) is a tri-linear map defined by
\begin{equation}
\label{GerPrDe}
\textstyle
\half
\left[m,m\right]
\left(
x_1,x_2,x_3
\right):=
m\left(m(x_1,x_2),x_3\right)-
m\left(x_1,m(x_2,x_3)\right).
\end{equation}
The condition $[m,m]=0$ is satisfied if and
only if $A=(E,m)$ is an associative algebra.

Consider the case where $E$ is split into a sum of 
two subspaces:
$$
E=V\oplus{}W
$$
and the bilinear map $m$ restricted to $V$ and $W$ 
satisfies the following conditions
\begin{equation}
\label{SplitM}
\begin{array}{ll}
m:V\times{}V\to{}V&
\hbox{is symmetric,}
\\[5pt]
m:V\times{}W\to{}W,&
m
\left|_{W\times{}V}\right.
\equiv0,
\\[5pt]
m:W\times{}W\to{}V&
\hbox{is skew-symmetric.}
\end{array}
\end{equation}
The following observation was done in \cite{LO}.

\begin{theorem}
\label{SkewThm}
The Lie antialgebra structure is a result of skew-symmetrization
of the equation $[m,m]=0$
in the elements of $W$.
\end{theorem}

The construction is as follows.
Let us use the notation: $x_i\in{}V$ and $y_j\in{}W$, and
consider the following terms
$$
\left[m,m\right]
\left(
x_1,x_2,x_3
\right),
\qquad
\left[m,m\right]
\left(
x_1,x_2,y
\right),
\qquad
\left[m,m\right]
\left(
x,y_1,y_2
\right),
\qquad\left[m,m\right]
\left(
y_1,y_2,y_3
\right),
$$
of the Gerstenhaber product
(i.e., the terms with $y$ never standing to the left of $x$).
Skew-symmetrize
the equation $[m,m]=0$ in $y$-variables, the above four terms 
then explicitly read:
\begin{eqnarray}
\label{First}
m\left(m(x_1,x_2),x_3\right)-
m\left(x_1,m(x_2,x_3)\right)&=&0,\\[6pt]
\label{Second}
m\left(m(x_1,x_2),y\right)-
m\left(x_1,m(x_2,y)\right)&=&0,\\[6pt]
\label{Third}
\textstyle
\half\,
m\left(m(x_1,y_1),y_2\right)-
\half\,m\left(m(x_1,y_2),y_1\right)
-m\left(x_1,m(y_1,y_2)\right)&=&0,\\[6pt]
\label{Fourth}
m\left(m(y_1,y_2),y_3\right)+\hbox{cycle}
&=&0
\end{eqnarray}
(note that $m\left(y_i,m(y_j,y_k)\right)$ in the last term
vanish by assumption (\ref{SplitM})).

Define the following operation on $E$:
\begin{equation}
\label{LeM}
\textstyle
\left]
x_1,x_2
\right[:=2\,m(x_1,x_2),
\qquad
\left]
x,y
\right[=
\left]
y,x
\right[:=m(x,y),
\qquad
\left]
y_1,y_2
\right[:=m(y_1,y_2).
\end{equation}
One then checks by a very simple computation that
the identities (\ref{First})--(\ref{Fourth})
are equivalent to the identities 
(\ref{AssCommT})--(\ref{Jack}) of Lie antialgebra.

\begin{rmk}
{\rm
The above construction suggests a generalization of the
notion of Lie antialgebra, cf. \cite{LO}.
More precisely, one can relax the symmetry condition 
for the map $m|_{V\times{}V}$ in (\ref{SplitM}).
In other words, one can assume in Definition \ref{GhostAl}
that the subalgebra $\fa_0$ in this case is associative but not necessarily commutative.
}
\end{rmk}

%%%%%%%%%%%%%%%%%%%%%%%%%%%%%%%%%%%%%%
\subsection{Relation to Lie superalgebras}
%%%%%%%%%%%%%%%%%%%%%%%%%%%%%%%%%%%%%%

There is a philosophical (or rather folklore) expression:
``a Lie superalgebra is but a square root of a Lie algebra''.
We argue that \textit{a Lie antialgebra is but a ``square root'' of a Lie superalgebra}.
More precisely, to an arbitrary Lie antialgebra,
we associate a Lie superalgebra
that plays an important r\^ole in the sequel.

Given a Lie antialgebra $\fa$,
consider the $\bbZ_2$-graded space
$
\fg_\fa=
\left(\fg_\fa\right)_0\oplus
\left(\fg_\fa\right)_1,$
where

\begin{enumerate}
\item
$
\left(\fg_\fa\right)_0=
\fa_1\odot_{\fa_0}\fa_1,
$
is the symmetric tensor square
${S^2\fa_1}$ over $\fa_0$,
i.e., consists of elements of the form
$(a\otimes{}b+b\otimes{}a)/_\sim$
where $a,b\in\fa_1$ and
where the equivalence is defined by 
$
]\a,a[\otimes{}b\sim{}a\otimes]\a,b[
$
for $\a\in\fa_0$;

\item
$\left(\fg_\fa\right)_1=\fa_1$.
\end{enumerate}

\noindent
The Lie bracket on $\fg_\fa$ is defined by
\begin{equation}
\label{CommKvad}
\begin{array}{rcl}
\left[
a\odot{}b\,,\,c\odot{}d
\right]&=&
\mathrm{Sym}_{(a,b),(c,d)}
\left(
\left]a,\left]b,c\right[\right[\odot{}d-
\left]c,\left]d,a\right[\right[\odot{}b
\right),\\[10pt]
\left[
a\odot{}b\,,\,c
\right]&=&
\left]a,\left]b,c\right[\right[+
\left]b,\left]a,c\right[\right[,\\[10pt]
\left[
a\,,\,b
\right]&=&
\left(a\otimes{}b+
b\otimes{}a\right)/_\sim,
\end{array}
\end{equation}
where $\mathrm{Sym}_{(a,b),(c,d)}$ is the symmetrisation in the pairs
$(a,b)$ and $(c,d)$.

\begin{theorem}
\label{SqThm}
The space $\fg_\fa$ endowed with the bracket
(\ref{CommKvad}) is a Lie superalgebra.
\end{theorem}

\noindent
For example,
$\fg_{\asl(2)}=\osp(1|2)$ and $\fg_{\cAK(1)}=\cK(1)$
coincide with the respective algebras of derivations.
(in general this is not the case).

%%%%%%%%%%%%%%%%%%%%%%%%%%%%%%%%%%%%%%%%%
%%%%%%%%%%%%%%%%%%%%%%%%%%%%%%%%%%%%%%%%%
\section{Representations of Lie antialgebras}
%%%%%%%%%%%%%%%%%%%%%%%%%%%%%%%%%%%%%%%%%
%%%%%%%%%%%%%%%%%%%%%%%%%%%%%%%%%%%%%%%%%

The notion of representation of a Lie antialgebras is
one of the most important.
It helps to better understand their nature and will hopefully be useful for applications.

%%%%%%%%%%%%%%%%%%%%%%%%%%%%%%%%%%%%%%%%%
\subsection{Definition and main example}
%%%%%%%%%%%%%%%%%%%%%%%%%%%%%%%%%%%%%%%%%

Consider a $\bbZ_2$-graded vector space  $V=V_0\oplus{}V_1$ and the following
``anticommutator'' on $\End(V)$:
\begin{equation}
\label{AntiCoCo}
\left]X,Y\right[:=
X\circ{}Y+(-1)^{p(X)p(Y)}\,
Y\circ{}X,
\end{equation}
where $p$ is the parity function on $\End(V)$
and  $X,Y\in\End(V)$ are homogeneous (purely even or purely odd)
elements.
The sign rule in (\ref{AntiCoCo}) is opposite to that of the
usual commutator.
Let us stress that the full space $\End(V)$ is \textit{not} a Lie antialgebra.

A representation of the Lie antialgebra  $\fa$ is 
an even linear map $\chi:\fa\to\End(V)$
such that
\begin{equation}\label{eqRep}
\left]\chi_x,\chi_y\right[=
\chi_{]x,y[},
\end{equation}
for all $x,y\in\fa$, cf. \cite{Ovs1}.

\begin{theorem}
\label{Reprthm}
Every representation of a Lie antialgebra $\fa$
extends to a representation of the corresponding Lie superalgebra $\fg_\fa$.
\end{theorem}

\noindent
The representation of the even elements of $\left(\fg_\fa\right)$ is given by
the operator
$$
X_{a\odot{}b}:=
\left[
X_a,X_b
\right]=
\chi_a\circ\chi_b+\chi_b\circ\chi_a,
\qquad
a,b\in\fa_1,
$$
which is the usual commutator of $\chi_a$ and $\chi_b$,
the odd elements are represented by $\chi$ itself.

\begin{rmk}
{\rm
The anticommutator (\ref{AntiCoCo}) coincides for odd $X,Y$ with
so-called twisted adjoint action, see \cite{Fra}.
One can say that Lie antialgebra structure arises from
this twisted adjoint action extended to the even part of $\End(V)$.
}
\end{rmk}

\begin{exe}
{\rm
Consider a $1|1$-dimensional space (i.g., $\bbC^{1|1},\bbR^{1|1}$ or $S^{1|1}$) with
coordinates $(x,\xi)$.
Fix an odd vector field
$$
D=
\frac{\partial}{\partial\xi}
+\xi\frac{\partial}{\partial{}x},
$$
sometimes called the ``SUSY-structure''.
It turns out that the map from $\cAK(1)$
to the space of vector fields proportional to $D$
\begin{equation}
\label{FRep}
\chi(\ell_i)=
x^{i+\half}\,D,
\qquad
\chi(e_n)=
\xi\,x^n\,D.
\end{equation}
is  a representation.

The corresponding representation of
the conformal Lie superalgebra $\cK(1)$ is given by the 
contact vector fields:
$$
X_h=
h(x,\xi)\,\frac{\partial}{\partial{}x}+
2\,D\left(h(x,\xi)\right)\,D,
$$
where $h(x,\xi)=h_0(x)+\xi\,h_1(x)$ is a (polynomial) function.
}
\end{exe}

%%%%%%%%%%%%%%%%%%%%%%%%%%%%%%%%%%%%%%%%%
\subsection{Representations of $\asl_2$}
%%%%%%%%%%%%%%%%%%%%%%%%%%%%%%%%%%%%%%%%%

Representations of the Lie antialgebra $\asl(2)$ were studied in \cite{MG}.
One of the most interesting results of this work is a close relation to so-called
ghost Casimir element of the universal enveloping algebra $U(\osp(1|2))$,
see \cite{Fra,Gor}.

A representation of $\asl(2)$ in a $\bbZ_2$-graded space $V=V_0\oplus{}V_1$
is given by one even operator $\E\in\End(V)_0$
and two odd operators $A,B\in\End(V)_1$ satisfying the relations
\begin{equation} 
\label{systemrep2}
\begin{array}{rcl}
AB -BA&=&\E\\ [5pt]
A\E + \E A&=&A\\ [5pt]
B\E + \E B&=&B\\ [5pt]
 \E^2&=&\E.
\end{array}
\end{equation}
These relations almost coincide with the canonical
Heisenberg relation $AB-BA=\id$, except that $\E\not=\id$
in our situation.
It is, indeed, shown in \cite{MG} that, up to equivalence,
$$
\E\left|_{V_0}\right.=0,
\qquad
\E\left|_{V_1}\right.=\id
$$
and deduced that $\asl(2)$ has no
non-trivial finite-dimensional
representations.

According to Theorem \ref{Reprthm}, every representation of
$\asl(2)$ is naturally a representation of the Lie superalgebra
$\osp(1|2)$.
The generators of this algebra are: $A,B$ together with
$$
E=A^2, 
\quad 
F=-B^2, 
\quad H=-(AB+BA).
$$
The following result determines the class of representations
of $\osp(1|2)$ that also support an $\asl(2)$-action.

\begin{theorem}
\label{NoOne}
There is a one-to-one correspondence between representations
of $\mathrm{asl}(2)$ and representations of
$\mathrm{osp}(1|2)$ such that 
\begin{equation}
\label{GhsAct}
\textstyle
\cc^2=\frac{1}{4}\,\id,
\end{equation}
where
$\cc$ is the action of the ghost Casimir element.
\end{theorem}

Recall that the \textit{ghost Casimir} element
$\cc\in{}U(\osp(1|2))$ is an invariants of the twisted adjoint action,
its action in an arbitrary representation is given by
$$
\textstyle
\cc=AB-BA-\frac{1}{2}\,\id.
$$
The above theorem can be deduced follows from the last relation (\ref{systemrep2}).

Another result of \cite{MG} is a classification of an interesting class of
irreducible infinite-dimensional representations of $\asl(2)$ called the weighted
representations.
This class of representations is related to the 
Harish-Chandra modules over $\osp(1|2)$, see \cite{BP}.
The following result is particularly nice.

\begin{theorem}
\label{OnlyOne}
The Lie antialgebra $\mathrm{asl}(2)$ has exactly one
highest weight representation and exactly one
lowest weight representation.
\end{theorem}

Many examples of representations of $\asl(2)$
are given by formula (\ref{FRep}) with $n=0$.

%%%%%%%%%%%%%%%%%%%%%%%%%%%%%%%%%%%%%%%%%
%%%%%%%%%%%%%%%%%%%%%%%%%%%%%%%%%%%%%%%%%
\section{Cohomology of Lie antialgebras}
%%%%%%%%%%%%%%%%%%%%%%%%%%%%%%%%%%%%%%%%%
%%%%%%%%%%%%%%%%%%%%%%%%%%%%%%%%%%%%%%%%%

Cohomology theory of Lie antialgebras was developed in \cite{LO}.
It is based on the classical ideas of Gerstenhaber \cite{Ger}
and Nijenhuis-Richardson \cite{NR} and uses graded Lie algebras
as the main tool.

%%%%%%%%%%%%%%%%%%%%%%%%%%%%%%%%%%%%%%%%%
\subsection{The notion of module}
%%%%%%%%%%%%%%%%%%%%%%%%%%%%%%%%%%%%%%%%%

Surprisingly, the notion of module over a Lie antialgebra \cite{Ovs1} is 
completely different from that of representation.

\begin{defi}
{\rm
Given a Lie antialgebra $\fa$,
a vector space $\cB$ is called an $\fa$-\textit{module} if the space
$\fa\oplus\cB$ is equipped with a Lie antialgebra structure satisfying the
following properties:
\begin{enumerate}
\item
the subspace $\fa\subset\fa\oplus\cB$ is a subalgebra isomorphic to the
initial Lie antialgebra, while $\cB\subset\fa\oplus\cB$ is an abelian
(trivial) subalgebra;
\item
the natural projection $\fa\oplus\cB\to\fa$ is a homomorphism
of Lie antialgebras.
\end{enumerate}
}
\end{defi}

It follows that for $a\in\fa$ and $b\in\cB$, one has
$]a,b[\in\cB$, so that one defines a linear map $\r:\fa\to\End(\cB)$
defined by
$$
\r_a(b)=]a,b[,
\qquad
a\in\fa,
\quad
b\in\cB.
$$
The Lie antialgebra structure on $\fa\oplus\cB$ is given by
$$
\left]
(a,b),\,(a',b')
\right[=
\left(
]a,a'[,\,\r_ab'+(-1)^{\s(a')\s(b)}\r_{a'}b
\right).
$$
It is called a semi-direct product and is denoted by
$\fa\ltimes\cB$.

\begin{exe}
\label{ModEx}
{\rm
(a)
The Lie antialgebra $\fa$ itself is an $\fa$-module
with $\r=\ad$.

(b) The dual space $\fa^*$ is an $\fa$-module, the map $\r$ being given by
$\r_a=(-1)^{p(a)}\ad^*_a$.
}
\end{exe}

\noindent
Note that neither $\ad$ nor $\ad^*$ is a representation.

%%%%%%%%%%%%%%%%%%%%%%%%%%%%%%%%%%%%%%%%%
\subsection{Combinatorial formula of the coboundary map}
%%%%%%%%%%%%%%%%%%%%%%%%%%%%%%%%%%%%%%%%%

Given a Lie antialgebra $\fa$ and an $\fa$-module $\cB$,
for all $p,q=0,1,2,\ldots$
we define the space, $C^{p,q}(\fa;\cB)$, of multi-linear
maps
\begin{equation}
\label{CochEq}
\vfi:
\underbrace{
\left(
\fa_0\otimes\cdots\otimes\fa_0
\right)
}_{p}
\otimes
\underbrace{
\left(
\fa_1\wedge\cdots\wedge\fa_1
\right)
}_{q}
\to\cB,
\end{equation}
skew-symmetric on $\fa_1$ and arbitrary on $\fa_0$.
Such a map is called a $(p,q)$-cochain.
We also consider the following space:
$$
C^k(\fa;\cB)=
\bigoplus_{q+p=k}C^{(q,p)}(\fa;\cB)
$$
that we call the space of $k$-cochains.

The coboundary operator $\d^k:C^k(\fa;\cB)\to{}C^{k+1}(\fa;\cB)$ is
defined as a sum of three operators:
\begin{equation}
\label{TheCobOpTot}
\d^k=\d^k_{1,0}+\d^k_{0,1}+\d^k_{-1,2},
\end{equation}
where $\d^k_{i,j}:C^{p,q}(\fa;\cB)\to{}C^{p+i,q+j}(\fa;\cB)$
for $p+q=k$.
The operators $\d^k_{1,0},\d^k_{0,1}$ and $\d^k_{-1,2}$
are given by the following explicit formul{\ae}.
If $q=0$, then $\d_{1,0}$ is the standard Hochschild differential.
If $q\neq0$, then
\begin{equation}
\label{TheCobOp10}
\begin{array}{l}
(\d^k_{1,0}\vfi)(x_1,\ldots,x_{p+1};\,y_1,\ldots,y_{q})
=\\[10pt]
\qquad
\qquad
\qquad
\displaystyle
m\left(
{x_1},\,\vfi(x_2,\ldots,x_{p+1};\,y_1,\ldots,y_{q})
\right)\\[10pt]
\qquad
\qquad
\qquad
\displaystyle
+\sum_{i=1}^{p}
(-1)^i\,\vfi(x_1,\ldots,x_{i-1},\,m(x_i,x_{i+1}),\,
x_{i+2},\ldots,x_{p+1};\,y_1,\ldots,y_{q})\\[14pt]
\qquad
\qquad
\qquad
\displaystyle
+
\frac{1}{q}
\sum_{j=1}^{q}(-1)^{p+j}\,
\vfi(x_1,\ldots,x_p;\,
m(x_{p+1},y_j),\,y_1,\ldots,\widehat{y_j},\ldots,y_{q}).
\end{array}
\end{equation}
\begin{equation}
\label{TheCobOp01}
\begin{array}{l}
(\d^k_{0,1}\vfi)(x_1,\ldots,x_{p};\,y_1,\ldots,y_{q+1})
=
\\[10pt]
\qquad
\qquad
\qquad
\displaystyle
C_k\,
\sum_{j=1}^{q+1}(-1)^{p+j}\,
m\left(
\vfi(x_1,\ldots,x_{p};\,
y_1,\ldots,\widehat{y_j},\ldots,y_{q+1}),\,{y_j}
\right)
\end{array}
\end{equation}
where
$$
C_k=
\left\{
\begin{array}{rl}
\frac{1}{q+1},&
\hbox{if $\vfi$ is with values in $W$ and $p\not=0$},\\[4pt]
\frac{2}{q+1},&
\hbox{if $\vfi$ is with values in $W$ and $p=0$},\\[4pt]
1,&
\hbox{if $\vfi$ is with values in $V$}.
\end{array}
\right.
$$
\begin{equation}
\label{TheCobOp-12}
\begin{array}{l}
(\d^k_{-1,2}f)(x_1,\ldots,x_{p-1};\,y_1,\ldots,y_{q+2})
=
\\[10pt]
\qquad
\displaystyle
\sum_{1\leq{}i<j\leq{}q+2}(-1)^{p+i+j+1}\,
\vfi(x_1,\ldots,x_{p-1},\,m(y_i,y_j)\,
;\,y_1,\ldots,\widehat{y_i},\ldots,\widehat{y_j},\ldots,y_{q+2}).
\end{array}
\end{equation}
(note that, to simplify the expression, we use the bilinear map $m$ instead
of the Lie antialgebra product $].,.[$, see formula (\ref{LeM})).

The following statement is a main result of \cite{LO}.
\begin{theorem}
The operator (\ref{TheCobOpTot}) satisfies the equation
$\d^{k+1}\circ\d^k=0$.
\end{theorem}

Unfortunately, the above formula is quite complicated.
Some of the terms are similar to the Hochschild differential
of associative algebras,
while some other terms are rather similar to the
Chevalley-Eilenberg differential of Lie algebras.

The space $\ker(\d^k)$ is called the space of $k$-cocycles
and the space $\mathrm{im}(\d^{k-1})$ the space of $k$-coboundaries.
We define the cohomology of a Lie antialgebra $\fa$ with
coefficients in an $\fa$-module $\cB$ in a usual way:
$$
H^k(\fa;\cB)=\ker(\d^k)/\mathrm{im}(\d^{k-1}).
$$
It has two subspaces, $H^k_\mathrm{ev}(\fa;\cB)$
and $H^k_\mathrm{odd}(\fa;\cB)$, of even and odd cohomology.

In the case, where $\cB$ is a
trivial module, the coboundary map $\d^k$ simplifies.
One has $\d_{0,1}=0$ and the following relations
$$
{\d_{1,0}}^2=0,
\qquad
{\d_{-1,2}}^2=0,
\qquad
\d_{1,0}\circ\d_{-1,2}+\d_{-1,2}\circ\d_{1,0}
=0.
$$
One therefore obtains a structure of bicomplex
with two commuting differentials.

%%%%%%%%%%%%%%%%%%%%%%%%%%%%%%%%%%%%%%%%%
\subsection{Lower degree cohomology, examples}
%%%%%%%%%%%%%%%%%%%%%%%%%%%%%%%%%%%%%%%%%

As in the usual Lie case, the cohomology classes of lower degree
have an algebraic interpretation.

\begin{proposition}
The space $H^1_\mathrm{ev}(\fa,\cB)$
describes extensions of
the $\fa$-module structure on $\cB$.
\end{proposition}

\noindent
Indeed, let $c:\fa\to\cB$ be a 1-cocycle on $\fa$ with values in $\cB$,
consider the space $\widetilde{\cB}=\cB\oplus\bbK$.
Define the following linear map
$\widetilde{\rho}:\fa\to \End(\widetilde{\cB})$
by
\begin{equation}
\label{Onedef}
\widetilde{\rho}_a\,(b,\l)=\left(\rho_a\,b+\l\,c(a),\,0\right).
\end{equation}

\begin{proposition}
The space $H^2_\mathrm{ev}(\fa,\cB)$
classifies the extensions of $\fa$ with coefficients in $\cB$;
trivial modules correspond to central extensions.
\end{proposition}

Almost no information is available about the cohomology of concrete Lie antialgebras.
The following example concerns the simplest possible case.

\begin{exe}
{\rm
Cohomology of $\asl(2)$ with trivial coefficients is trivial.
}
\end{exe}

\noindent
It is conjectured in \cite{LO} that the cohomology with trivial coefficients
of a Lie antialgebra $\fa$ alway vanishes provided the subalgebra
$\fa_0$ contains the unity.

In the infinite-dimensional case, only a few examples of
non-trivial cohomology classes are known \cite{LO}.

\begin{theorem}
The linear map $\g:\cAK(1)\to\cAK(1)^*$ given by
$$
\textstyle
\g(e_n)=-n\,e_{-n}^*,
\qquad
\g(\ell_i)=\left(\ell^2-\frac{1}{4}\right)\ell_{-n}^*,
$$
is a non-trivial 1-cocycle on $\cAK(1)$.
\end{theorem}

We understand the above cocycle as analog of the famous Gelfand-Fuchs cocycle
related to the Virasoro algebra.
Another non-trivial cohomology class similar to the
Godbillon-Vey class is also constructed in \cite{LO}.

It would be interesting to investigate applications of Lie antialgebra
cohomology in geometry and topology.

\bigskip

\noindent \textbf{Acknowledgments}.
I am grateful to P. Lecomte and S. Morier-Genoud
for enlightening discussions.
This paper was conceived during the XXVII Workshop
``Geometrical Methods in Mathematical Physics'',
I am pleased to thank the organizers for this wonderful conference.

%%%%%%%%%%%%%%%%%%%%%%%%%%%%%%%%%%%%%%%%%%%%%

CNRS,

Institut Camille Jordan,

Universit\'e Lyon 1,

Villeurbanne Cedex, F-69622,

France;

ovsienko@math.univ-lyon1.fr

\end{document}